\documentclass[usegraphicx]{mn2e}

\usepackage{amssymb}

\input{epsf}

\begin{document}

\title{Cluster infall in the concordance $\Lambda$-CDM model.}
\author[M. C. Pivato, N. D. Padilla \& D. G. Lambas]
{Maximiliano C. Pivato$^1$, Nelson D. Padilla $^{1,2}$ and Diego G. Lambas$^{1,3}$ \\
$^1$ Grupo de Investigaciones en Astronom\'{\i}a Te\'{o}rica y Experimental 
 (IATE), Observatorio Astron\'{o}mico C\'{o}rdoba, UNC, Argentina.\\
$^2$ Departamento de Astronom\'{\i}a y Astrof\'{\i}sica, Pontificia 
Universidad Cat\'olica, V. Mackenna 4860, Santiago 22, Chile.\\
$^3$ Consejo Nacional de Investigaciones Cient\'{\i}ficas y Tecnol\'ogicas 
(CONICET), Argentina.\\}

\date{Accepted ???? Month ??. Received ???? Month ??}
\maketitle

\begin{abstract}
We perform statistical analyses of the infall of dark-matter
onto clusters in numerical simulations within the concordance ${\Lambda}$CDM
model. By studying the infall profile around clusters of different 
mass, we find a linear relation between the mass and maximum infall velocities
which reach $\sim 900$ km/s for the most massive groups. 
The maximum infall velocity and the group mass follow a suitable power law 
fit of the form, $V_{inf}^{max} = (M/m_{0})^{\gamma}$. By comparing the 
measured infall velocity to the linear infall model with an exponential 
cutoff introduced by Croft et al., we find that the best agreement is 
obtained for a critical overdensity $\delta_{c} = 45$. 
We study the dependence of the direction of infall with respect
to the cluster centres, and find that in the case of massive groups,
the maximum alignment occurs at scales $r \sim 6h^{-1}$ Mpc. 
We obtain a logarithmic power-law relation between the average 
{\it infall angle} and the group mass. 
We also study the dependence of the results on the local dark-matter 
density, finding a remarkable difference in the dynamical behaviour of 
low- and high-density particles. 
\end{abstract}


\section{Introduction}

The large-scale velocity field provides a significant source of 
information on the distribution of mass fluctuations in the Universe. 
Moreover, the relation between peculiar velocities and mass is simple, 
providing a direct indication of the distribution of mass independent 
of assumptions about the bias factor of different galaxy types. 

Large scales flows have been 
addressed both theoretically (Reg\"{o}s \& Geller 1989, Bothun et al. 1990, 
Hiotelis 2001) and observationally (Dekel 1999, Cecarrelli et al. 2005, for 
a review see Strauss \& Willick 1995). 
On the other hand, the dynamical behavior of objects near
regions of high density contrast can be described by the spherical 
infall model (Reg\"{o}s \& Geller 1989) with a collapsing streaming motion 
whose amplitude depends only on the distance to the local density maximum 
(Diaferio \& Geller 1997). The predictions from the spherical infall model 
are often used for cluster mass estimates 
({\it{e.g.}} Diaferio 1999). 
Gunn (1978) and Peebles (1976, 1980) derived a linear approximation to the 
infall velocity induced by an isotropic mass concentration described by
$\delta(r)=\rho(r)/\rho_{b} - 1$, where $\rho(r)$ is the 
average density inside a radius $r$ and $\rho_{b}$ is the background 
density. The infall velocity $V_{inf}$ is then given by
\begin{equation}
V_{inf}^{lin} = -\frac{1}{3}H_{0}\Omega_{0}^{0.6}r\delta(r),
\label{vlin}
\end{equation} 
where $H_{0}$ is the Hubble constant and $\Omega_{0}$ is the 
cosmological density parameter. Yahil (1985) provides a non-linear 
approximation to the exact solution of the infall velocity:
\begin{equation}
V_{inf}^{non-lin} = -\frac{1}{3}H_{0}\Omega_{0}^{0.6}r
\frac{\delta(r)}{[1 + \delta(r)]^{0.25}};
\label{yahil}
\end{equation} 
however, this equation is not accompanied by a published derivation. 
Reg\"{o}s \& Geller (1989), assume that clusters are spherically symmetric and
obtain a solution for the non-linear infall velocity of the system as an 
expansion series, which is accurate before orbit crossing takes place and for 
$\delta < 1-2$.  They also quote that this solution is inadequate for 
large density enhancements.  Therefore, neither their calculations 
nor Eqs. \ref{vlin} or \ref{yahil}, are expected to hold in or 
near cluster centres, or more generally 
in virialised regions.   In order to overcome this problem,
Croft, Dalton \& Efstathiou (1999) -hereafter, CDE-  
choose to truncate the expression for $V_{inf}^{non-lin}$ with an exponential 
cutoff $e^{-\delta(r)/\delta_{c}}$ with $\delta_{c}=50$, which roughly 
approximates the effect of the rapid decrease of infall velocities at small 
$r$ due to virialised motions (see figure 7 in CDE).  When adding the 
exponential cutoff, the expression for the linear infall velocity 
given by Eq. \ref{vlin} becomes:
\begin{equation}
V_{inf}^{lin} = -\frac{1}{3}H_{0}\Omega_{0}^{0.6}r\delta(r)
               e^{\delta(r)/\delta_{c}};
\label{cvlin}
\end{equation}
for the non-linear infall velocity (Eq. \ref{yahil}) this is:
\begin{equation}
V_{inf}^{non-lin} = -\frac{1}{3}H_{0}\Omega_{0}^{0.6}r
\frac{\delta(r)e^{-\delta(r)/\delta_{c}}}{[1 + \delta(r)]^{0.25}}.
\label{cyahil}
\end{equation}
CDE claim to obtain a better fit to their simulations using this
equation, but given the arbitrary nature of this 
truncation, they still restrict their analysis to the outer regions of haloes 
($r > 2.5h^{-1}$ Mpc).

In this paper, we explore the characteristics of streaming motions in the
regions surrounding clusters using large, high resolution VIRGO N-body simulations,
and compare these results to predictions from linear theory and the
spherical infall model.  We will also study the alignments between
the velocity field and the positions of haloes in the numerical simulation,
to determine the extent to which a radial approximation to the velocity 
field around
high-density peaks is valid.  In a related subject, we will provide a further
quantification of velocity alignments by measuring the fractions of 
particles flowing towards haloes with respect to those staying at roughly 
fixed gravitational potentials, and we will compare it to the fraction of 
particles escaping the potential wells of haloes. In all these analyses 
we will also study the impact of local density on the infall and 
alignment of motion towards clusters.

This paper is organised as follows: Section 2 contains a brief description
of the numerical simulations used in this work;
Section 3 describes the statistical analyses and the resulting
dependencies of the infall pattern on cluster mass and local density.  
Finally, Section 4 contains a summary of the conclusions drawn from this work.

\section{Numerical Simulations}

The numerical simulation used in this work was carried out by the Virgo 
Consortium. In particular, we use the publicly available 
Very Large Simulations, VLS (Yoshida et al. 2001), characterised by
a ${\Lambda}CDM$ model, $512^3$ particles in a 
cosmological box of $480h^{-1}$ Mpc on a side, and a mass resolution of
$6.86\times10^{10}h^{-1}M_{\odot}$.  The cosmology is characterised by a
matter density parameter $\Omega_0=0.3$, cosmological constant 
$\Omega_{\Lambda}=0.7$ 
and expansion rate at the present time $H_0=100h$ Mpc km/s where $h=0.7$.
The Cold Dark Matter (CDM) initial power spectrum was 
computed using the $CMBFAST$ software (Seljak \& Zaldarriaga 1996), and is 
normalised to the abundance of galaxy clusters at $z=0$ so that 
$\sigma_8=0.9$. Dark-matter haloes are identified in the 
simulation using a standard {\it{friends-of-friends}} algorithm 
(Davis et al. 1985) using a linking 
length $l$, given by $l=0.17\bar{n}^{-1/3}$, where $\bar{n}$ is the mean 
number density of particles; this linking length corresponds to an 
overdensity $\delta\rho/\rho \simeq 200$, which 
characterises the virialized halo and separates it from the sorrounding 
infall material (Cole \& Lacey 1995).
This large simulation contains a large number of halos spanning 
a wide range of masses. The criteria followed in the identification of 
dark-matter haloes leads to $\sim 720,000$ identified systems with masses 
ranging from  $6.8\times10^{11}h^{-1}M_{\odot}$ 
(10 particles) to $2.35\times10^{15}h^{-1}M_{\odot}$ (more than $20,000$
particles).

\begin{figure}
       \includegraphics[width=0.47\textwidth]{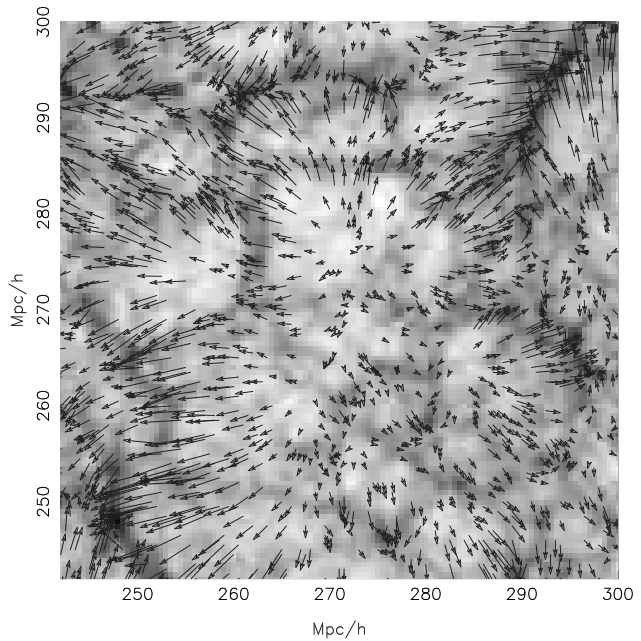}
       \caption{Infall pattern in a slice, $60h^{-1}$ Mpc a side and
$10h^{-1}$ Mpc thick, of the VLS simulation box. The particle density is 
plotted in a logarithmic gray-scale, smoothed using a top-hat function  
($R=1h^{-1}$ Mpc). The black solid arrows show the peculiar velocity field.}
       \label{fig0}
\end{figure}

In Figure \ref{fig0} we plot a slice of the VLS simulations where
the distribution of particles is shown in a logarithmic gray-scale 
density plot. The dark matter density was smoothed using a $R=1h^{-1}$ Mpc 
top-hat filter and corresponds to a slice of $60h^{-1}$ Mpc a side and 
$10h^{-1}$ Mpc deep. The black arrows indicate the projected 
velocity field of $20\%$ of the particles in the slice. It can be 
clearly appreciated the systematic infall flow onto dense regions, as 
well as outflow patterns from the lower density regions.

\begin{figure}
       \includegraphics[width=0.47\textwidth]{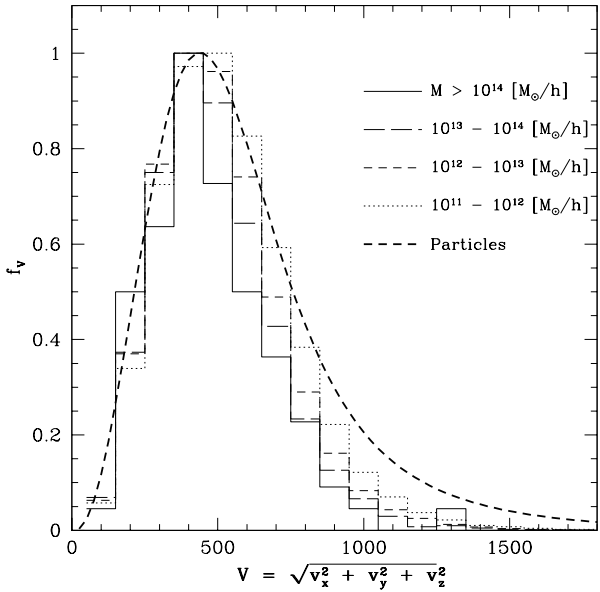}
       \caption{Normalised distribution function, $f_v$, of the velocity
module $\textbf{V}=\sqrt{v_x^2 + v_y^2 + v_z^2}$ for particles (thick short 
dashed line) and groups with masses in the ranges 
$10^{11}h^{-1}M_{\odot} - 10^{12}h^{-1}M_{\odot}$ (dotted line), 
$10^{12}h^{-1}M_{\odot} - 10^{13}h^{-1}M_{\odot}$ (short dashed line), 
$10^{13}h^{-1}M_{\odot} - 10^{14}h^{-1}M_{\odot}$ (long dashed) and more than 
$10^{14}$ $h^{-1}M_{\odot}$ (solid line).}
       \label{fig1}
\end{figure}

We calculate the halo peculiar velocity by averaging the individual group
member velocities.
In Figure \ref{fig1} we show the normalised distribution function of the 
peculiar velocity of halos with masses between 
$10^{11}h^{-1}M_{\odot} - 10^{12}h^{-1}M_{\odot}$ (dotted line), 
$10^{12}h^{-1}M_{\odot} - 10^{13}h^{-1}M_{\odot}$ (short dashed line), 
$10^{13}h^{-1}M_{\odot} - 10^{14}h^{-1}M_{\odot}$ (long dashed line) and 
more than $10^{14}h^{-1}M_{\odot}$ (solid line); and 
the distribution function of peculiar velocities of the dark matter particles
(thick short-dashed line).
As can be seen, even the most massive groups have similar mean-velocities 
than dark-matter particles (Einasto et al. 2005, Shaw et al. 2005).  
However, only dark-matter particles show a 
high velocity tail (${\bf{V}} > 1500$ km/s), whereas groups rarely reach
peculiar velocity values above ${\bf{V}} = 1000$ km/s (Shaw et al. 2005).

\section{Velocity Field Analysis}

In this section we study the statistical properties of the velocity 
field in the neighborhoods of dark-matter haloes, up to scales of 
$40h^{-1}$ Mpc. We begin by analysing the mean infall velocity of particles 
onto groups, and the distribution of the peculiar velocity orientation 
of the particles in the outskirts of the groups (Subsection \ref{infv}). 
In Subsection \ref{infa} we study the dependence of the results 
on the local dark-matter density. We perform our analysis for four 
diferent group mass samples (see Table \ref{tab:samp}).

All our analises are performed using individual particles in the simulation
as opposed to the recent use of substructure identified in simulations
in several works (see for instance Benson, 2005, and Wang et al., 2005).
We consider that our studies should not change importantly if we were
to repeat the following analyses using substructure since we are particularly
interested in regions outside the halo centres.  At such distances, we
expect the average velocities of individual particles and substructre
to be similar.  This is opposed to the clear differences inside dark-matter
haloes, where dynamical friction would slow down the center of mass of
a sub-halo.

\begin{table}
    \caption{Description of our samples of dark-matter haloes
 corresponding to different ranges of mass and the number of groups 
 identified.}
    \begin{center}
        \begin{tabular}{ccc}
            \hline\hline
Sample & Mass interval $[h^{-1}M_{\odot}]$ & Number of groups\\
            \hline\hline
S1  & $3.4\times10^{12} - 6.8\times10^{12}$ &  $73509$\\
S2  & $6.8\times10^{12} - 6.8\times10^{13}$ &  $72465$\\
S3  & $6.8\times10^{13} - 6.8\times10^{14}$ &  $6097$\\
S4  & $>6.8\times10^{14}$ & $92$\\
            \hline\hline
        \end{tabular}
    \end{center}
    \label{tab:samp}
\end{table}

\subsection{The average velocity infall} \label{infv}

\begin{figure}
       \includegraphics[width=0.47\textwidth]{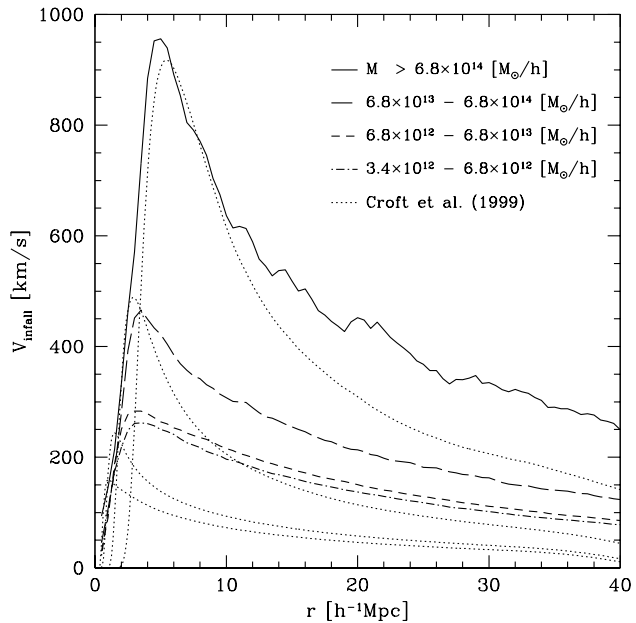}
       \caption{Average infall velocity 
onto groups in samples S1, S2, S3 and S4 (dot dashed, short dashed, 
long dashed and solid lines, respectively) as function of scale. Dotted 
lines represent the results from the linear fit with an exponential 
cutoff (Eq. \ref{cvlin}) with $\delta_{c}=50$.} 
       \label{fig2}
\end{figure}

We start our analysis of the velocity field by computing the 
group-averaged infall velocity as follows: we select groups in a 
specific mass range, and calculate the mean particle infall velocity
as a function of distance to the group centres.
In Figure \ref{fig2} we plot the mean infall velocity profile for the four
subsamples S1, S2, S3 and S4 (dotted, short-dashed, long-dashed and solid 
line respectively) described in Table \ref{tab:samp}.
As it can be seen, the maximum infall velocity, $V_{inf}^{max}$, 
increases with the group mass, with values $V_{inf}^{max} \sim 250$ km/s 
for groups in sample S1, and $V_{inf}^{max} = 950$ km/s for those 
in S4, the most massive groups in our analysis. The rapid decrease of infall 
velocities at small scales indicates the boundary 
of the virialised region. It can also be seen that the scale 
of the maximum infall velocity correlates well with 
group mass, a somewhat expected effect given the correlation between 
mass and virial radius.
We notice a more rapid decrease of the infall pattern for the most massive 
groups compared to the gentle decline shown by low mass systems.
We compare the actual infall velocity measured in the simulations with 
the prediction from the CDE model. 
In order to do this, we calculate the averaged density enhancement inside a 
sphere of radius $r$ centered in each group and derive the linear 
infall velocity using an exponential cutoff given by Eq.
\ref{cvlin} for the same mass intervals 
described in Table \ref{tab:samp}. The results are plotted in Figure 
\ref{fig2}, where it can be appreciated that although this  model provides 
a reasonable agreement to the infall around high mass haloes, there are still large 
discrepancies at small and large scales, specially for the lower mass ranges 
explored here.  

\begin{figure}
       \includegraphics[width=0.47\textwidth]{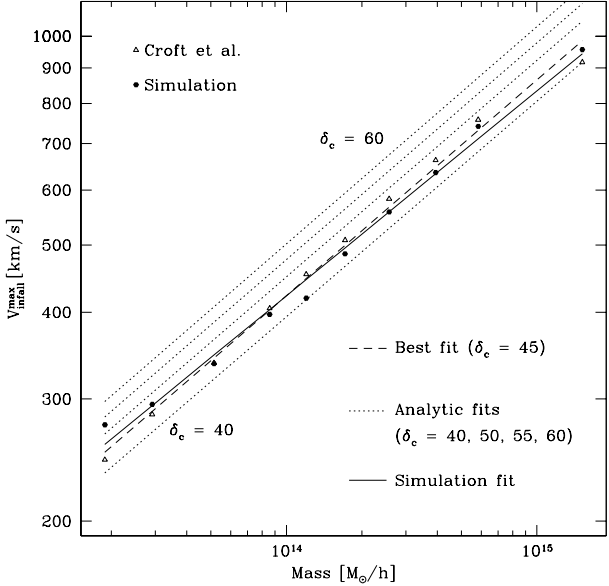}
       \caption{Maximum infall velocity as a function of group mass 
       measured in the simulation (black dots) and its corresponding power-law fit 
       (solid line). The open triangles show the same relation as obtained 
       from the maxima in Eq. \ref{cvlin}.  We also show for comparison the
       results from Eq. \ref{vmax}, assuming different values of 
       $\delta_{c} = 40, 45, 50, 55, 60$ (dotted lines).  The best-fit to the
       simulation results is given by the dashed line, which corresponds to 
       $\delta_{c}=45$.}
       \label{fig3}
\end{figure}

The results shown in Figure \ref{fig2} indicate a tight 
relation between $V_{inf}^{max}$ and group mass in both, the patterns 
measured in the simulations, and the results from 
Eq. \ref{cvlin}. In order to address the infall correlation with 
group mass we measure the mean infall profile for different mass ranges, 
and compute $V_{inf}^{max}$. In Figure \ref{fig3}, we show the maximum 
mean infall velocity as a function of group mass in the 
simulation (solid circles), and from Eq. \ref{cvlin} (open triangles). 

A suitable power-law fit to the simulation results is:
\begin{equation}
    V_{inf}^{max} = \left(\frac{M}{m_0}\right)^{\gamma},
    \label{fit}
\end{equation}
where the parameter values are
$m_{0} = 1.29\times10^{5}$, and $\gamma = 0.29 $. Using the linear fit 
with the exponential cutoff given by Eq. \ref{cvlin}, it is possible 
to infer a relation between the scale of the maximum infall velocity and 
the halo mass:
\begin{equation}
    r_{inf}^{max} = \left(\frac{3A}{8\pi\rho_{m}}\right)^{1/3}M^{1/3}.
    \label{rmax}
\end{equation}
where the factor $A$ is defined as function of $\delta_{c}$ as
$A=\sqrt{(2+3/\delta_{c})^2 + 12/\delta_{c}} - 2 - 3/\delta_{c}$, 
with $A>0$ for any given positive value of $\delta_c$. Replacing $r$ by
$r_{inf}^{max}$ in Eq. \ref{cvlin}, we find the expression for 
$V_{inf}^{max}$:
\begin{equation}
    V_{inf}(r_{max}) =
    \frac{(9\pi)^{-1/3}}{2}\frac{H_{0}\Omega_{0}^{0.6}}{\rho_{m}}
    \left(\frac{2-A}{A^{2/3}}\right)e^{ -\frac{2-A}{A\delta_{c}}}
    M^{1/3}.
    \label{vmax}
\end{equation}
By setting $H_{0}=70$, $\Omega_{m}=0.3$ and $\delta_{c}=50$, we 
obtain $V_{inf}(r_{max})\simeq M^{1/3}/2.25\times10^4$, which is 
very similar to our power-law fit in Eq. \ref{fit}. 
We now calculate $V_{inf}^{max}(M)$ from Eq. \ref{vmax} using 
different values of $\delta_{c}= 40$, $45$, $50$, $55$ and $60$, 
and plot the results in Figure \ref{fig3} (dotted lines). 
As it can be seen in this figure, the best fit to the simulations is achieved
when using $\delta_{c} \sim 45$ (dashed line). The power-law behaviour of
$V_{inf}^{max}(M)$  brakes down at low mass halos, when 
$M < 10^{13}h^{-1}M_{\odot}$.  
At this point,
the maximum infall velocity does not change 
considerably with halo mass, as can be seen in the infall patterns for 
samples S1 and S2 in figure \ref{fig2}.  
The reason behind this may
be related to the flattening of the bias between dark-matter haloes
and mass $b(M)$, which occurs at similar mass scales; namely, the
clustering around groups of masses $M < 10^{13}h^{-1}M_{\odot}$ can
be considered to be almost independent of halo mass.

\begin{figure}
       \includegraphics[width=0.47\textwidth]{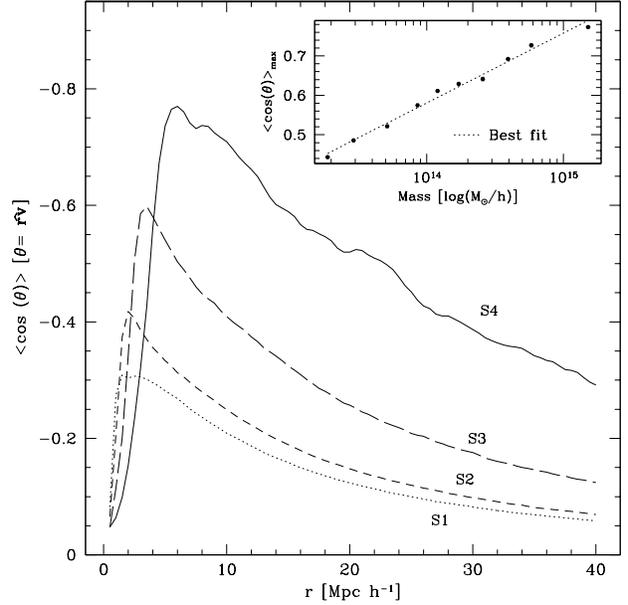}
       \caption{Average cosine of the angle between the
cluster-centric position and peculiar velocity of dark-matter particles 
around haloes in the simulation.  Groups are divided into samples 
spanning different mass ranges indicated in Table \ref{tab:samp} 
(samples S1, S2, S3 and S4, in dotted, short dashed, long dashed 
and solid lines, respectively). The inset shows the relation between 
$\left<cos(\theta)\right>_{max}$ and group mass. The dotted line 
corresponds the best log-linear fit to this relation, 
$cos(\theta)_{max} = {\gamma}log(M/C_{0})$, where $\gamma = 0.17$ 
and $C_{0} = 5.18\times10^10$. }
       \label{fig5}
\end{figure}

In order to characterise the alignments of the infall of dark-matter
in the outskirts of haloes, we centre our origin on
each halo, and measure the angle $\theta$ between the particle position 
vector and its peculiar velocity. In order to detect variations with the 
halo mass, we analyse subsamples S1, S2, S3 and S4 described in Table 
\ref{tab:samp}. Figure \ref{fig5} shows the results from averaging 
$\left<cos(\theta)\right>$ over each sample of haloes (dotted, 
short dashed, long dashed and solid lines for samples S1, S2, S3 and S4, 
respectively).

Dark-matter particles infalling towards S4 groups show the
maximum alignment, $cos(\theta)_{max} \simeq - 0.77$ 
at $r \simeq 5.5h^{-1}$ Mpc. In contrast, the peculiar velocity of  
particles in the outskirts of groups in sample S1 show the lowest
alignment,  with $cos(\theta)_{max} \simeq - 0.3$ at $r \simeq 1h^{-1}$ Mpc. 
As can be seen, we obtain a maximum `coherence' for the alignment in a 
relatively small range of scales: 
$1h^{-1}$ Mpc $\lesssim r \lesssim 6h^{-1}$ Mpc. The velocity field of 
the particles around the most massive groups (sample S4) show a significant 
alignment at large scales ($r > 30h^{-1}$ Mpc). However, for less massive 
systems, $\left<cos(\theta)\right>$ is significantly reduced. 
It can also be appreciated that the maximum alignment between the
cluster-centric position and the peculiar velocity of the particles is
a function of group mass. We show this relation in the inset of 
Figure \ref{fig5} for different masses (same samples as those in Figure 
\ref{fig3}). An empirical logarithmic power-law fit,
\begin{equation}
cos(\theta)_{max}(M) = {\gamma}log(M/C_{0}),
\end{equation}
with parameters $\gamma = 0.17$ and $C_{0} = 5.18\times10^{10}$,
gives a very good agreement to the simulation results, 
as can be seen in dotted line in the inset 
of Figure \ref{fig5}.

\begin{figure}
       \includegraphics[width=0.47\textwidth]{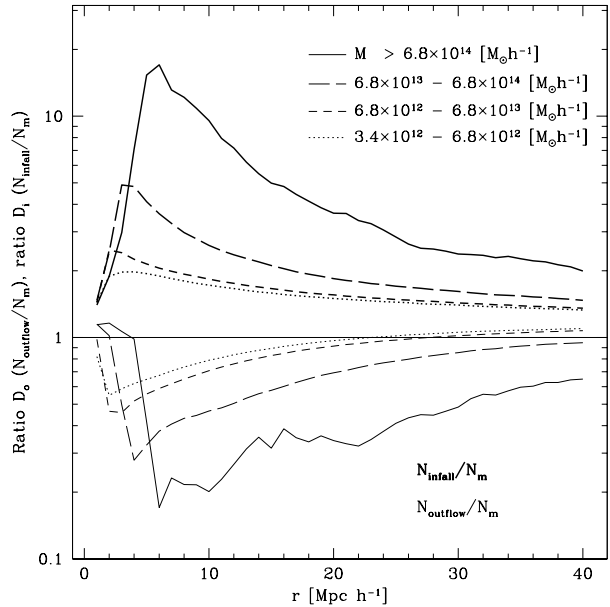}
       \caption{The average ratio between the number of particles with 
$1/3 < cos(\theta) <~ 1$ ($N_{outflow}$) and $-1/3 < cos(\theta) < 1/3$ 
($N_{m}$), referred to as $D_i$ (black); and the ratio between the number of 
particles with $-1 < cos(\theta) < -1/3$ ($N_{infall}$) and $N_{m}$, 
refereed to as $D_o$ (red). These ratios are plotted for the different 
mass ranges used in previous plots: 
$3.45\times10^{11} - 6.8\times10^{12} h^{-1}M_{\odot}$ (dotted line), 
$6.8\times10^{12} - 6.8\times10^{13} h^{-1}M_{\odot}$ (short dashed line), 
$6.8\times10^{13} - 6.8\times10^{14} h^{-1}M_{\odot}$ (long dashed line) and 
$M > 6.8\times10^{14} h^{-1}M_{\odot}$ (solid line).}
       \label{fig7}
\end{figure}

We perform a further study of the {\it{infall-angle}} in the outskirts of 
dark-matter haloes by comparing the number of particles in each shell of radius 
$r$ with peculiar velocities pointing outwards from the group center, to 
the number pointing towards the group center. 
We measure $N_{i}(r)$, $N_{m}(r)$ and 
$N_{o}(r)$, which indicate the number of particles characterised by 
$1/3 \lesssim cos(\theta)$ (infall), 
$-1/3 < cos(\theta) < 1/3$ (intermediate) and 
$cos(\theta) \lesssim -1/3$ (outflow), respectively, at a distance $r$ 
from the group centre.  We average across our sample, and calculate the ratios
$D_{i}(r) = <( N_{i}(r)/N_{m}(r) )>$ and $D_{o}(r) = <( N_{o}(r)/N_{m}(r) )$.
Figure \ref{fig7} show the results for $D_{i}(r)$
(thick lines) and $D_{o}(r)$ (thin lines) for the samples S1, S2, S3 and S4 
(dotted, short-dashed, long-dashed and solid lines, respectively).
As can be seen, samples S1 and S2 show $D_{o} > 1$ at $r > 30h^{-1}$ Mpc,
indicating that the large density inhomogeneities marked by
the halo centres are still correlated with the distribution of 
mass at such large distances.

The fact that the maximum of $D_i$ and minimum of $D_o$ coincide
with one another, as well as with the maximum alignment as measured by 
$cos(\theta)$, is to be expected since the maximum alignment occurs when the
fraction of infalling particles reaches its maximum 
(compare figure \ref{fig5} and \ref{fig7}). We also notice 
that these maxima occur at larger distances for higher mass haloes, 
consistent with the maximum infall velocities occurring further away 
from the most massive haloes.
We notice that the maximum infall velocity occurs $\sim 30\%$ closer to the 
group than the maximum velocity alignment.  This suggests that 
alignments are a better choice than infall velocities for
the determination of the distance where the halo virialisation
breaks the trend of increasing infall and velocity alignments toward
the halo centres.
In general, our results are consistent with high mass 
haloes being more relaxed systems with isotropic velocity distributions.  
%

\subsection{The infall dependence on local density.}
\label{infa}

In this subsection, we analyse the dependence of the infall pattern, the 
{\it{infall-angle}} and the parameters $D_{i}$ and $D_{o}$ on the 
local density of the dark-matter particle being considered in the 
analysis.  We define a local number density, $\rho_{n}$, as 
the number of particles in cubic cells of side $l_{c}=3h^{-1}$ Mpc (note this 
is not a proper density definition, just a simple parameter to characterise the 
environment of a dark-matter particle), and a mean numerical density, 
$\bar\rho_{n}$, as the mean number of particles expected in cells of 
volume $V_{c} = l_{c}^{3}$ (in this particular numerical simulation, and 
for the chosen value of $l_{c}$, the mean numerical density is 
$\bar{\rho_{n}} = 33$).  All particles within a cubical cell are assigned
the same density and we divide our analysis according to
the following ranges of local density:
{\it{high}}, {\it{intermediate}} and {\it{low densities}} 
defined as $\rho_{n}/\bar{\rho_{n}} > 2$, 
$1/2 < \rho_{n}/\bar{\rho_{n}} < 2$ and $\rho_{n}/\bar{\rho_{n}} < 1/2$,
respectively.

\begin{figure}
       \includegraphics[width=0.47\textwidth]{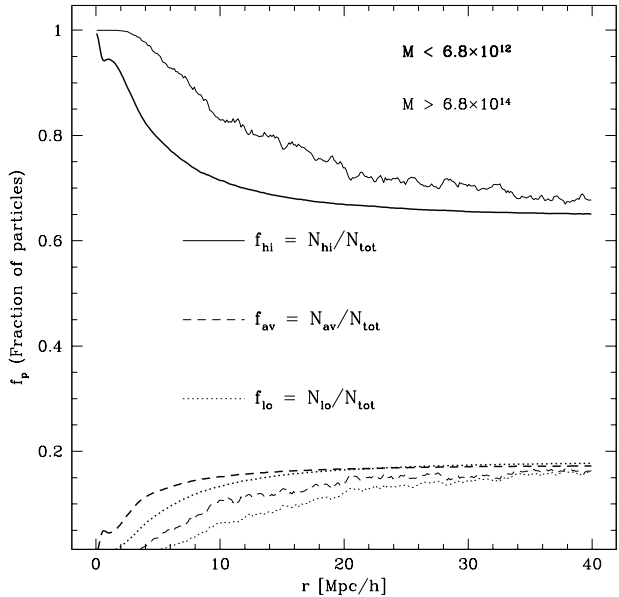}
       \caption{
       Fraction of particles in high, intermediate and low density regions ,
 $f_{hi}$
       (solid lines),  $f_{av}$ (dashed lines)
       and $f_{lo}$ (dotted lines), respectively as a function of
       scale. The thick lines correspond to the most massive groups (sample S4)
       and the thin lines to the least massive ones (sample S1).
       }
       \label{figend}
\end{figure}

In order to test the possibility of a further dependence of the infall 
pattern on mass and local density, we analysed the fraction of
particles laying in diferent density ranges as function of  scale. The
fraction of particles in high, intermediate and low density regions are 
calculated using $f_{hi} = N_{hi}/N_{tot}$, $f_{av} = N_{av}/N_{tot}$
and $f_{lo} = N_{lo}/N_{tot}$, where  $N_{hi}$, $N_{av}$ and $N_{lo}$
are the number of particles in high, intermediate and low density regions 
respectively, and $N_{tot} = N_{hi} + N_{av} + N_{lo}$ is the total number 
of particles within each distance interval. In figure \ref{figend} we plot with thick lines the fraction of particles $f_{hi}$ (solid lines), 
$f_{av}$ (dashed lines) and $f_{lo}$ (dotted lines) for the S4 
$M > 6.8\times10^{14}$ sample, and with thin lines the corresponding 
fraction of particles for sample S1 ($M < 6.8\times10^{12}$). In this 
figure, we can appreciate  that the infall patterns corresponding to $f_{hi}$
for the two mass range sample differ in less than $20\%$ in scales 
$r < 20 Mpc/h$ while the patterns of $f_{av}$ and $f_{lo}$ are similar within
 $10\%$ in the same scales.

\begin{figure}
       \includegraphics[width=0.47\textwidth]{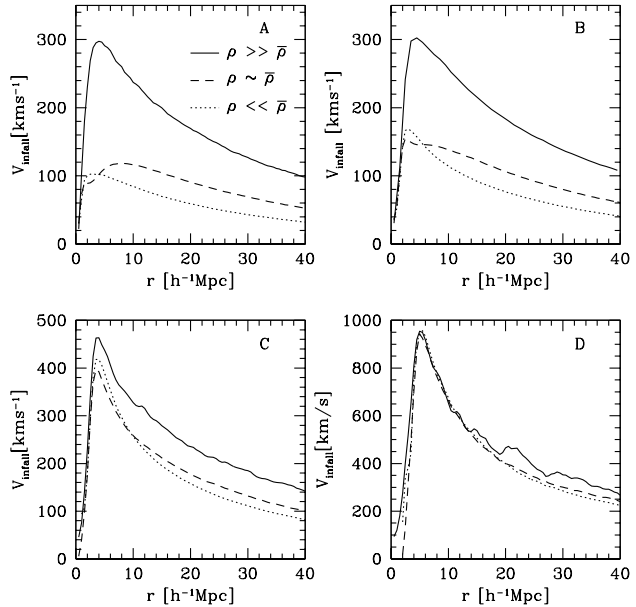}
       \caption{Mean infall velocity as a
function of scale.  Halo samples correspond to S1, S2, S3 and S4, and are
defined in Table \ref{tab:samp} (panels A, B, C and D). The different 
lines show the average infall patterns of particles in 
regions with density $\rho >> \bar{\rho}$ (solid line), 
$\rho \sim \bar{\rho}$ (dashed line) and $\rho << \bar{\rho}$ 
(dotted line).
}
       \label{fig4}
\end{figure}

In Figure \ref{fig4} we plot the average infall velocity
of particles with $\rho/\bar{\rho} > 2$ (solid lines), 
$\rho/\bar{\rho} \sim 1$ (dashed lines) and $\rho/\bar{\rho} < 1/2$ 
(dotted lines) as a function of the scale, for the four group samples S1, S2, 
S3 and S4 (panels A, B, C and D respectively).  It can be noticed that  
more massive haloes tend to show less differences in the infall patterns 
of high and low  density particles (almost no difference in panel D), whereas low 
mass groups (panel A) show a difference between maximum infall of high 
and low density particles of $\sim 200$ km/s (more than $60\%$ of the 
maximum infall for the high density particles).  
We provide fits to
the $V_{inf}^{max}$ vs. mass relation for the different ranges of
local density in table \ref{tab:samp2}, where it can be seen that both the
slope, $\gamma$, and $m_0$ increase for decreasing local density
values.  Note that the global fit is similar to that of $\rho_n > 2\bar{\rho_n}$,
which indicates that the signal from high density regions dominates the
overall infall velocity.
\begin{table}
       \caption{Values obtained for the parameters $\gamma$ and $m_0$ in the fit
      of $V_{inf}^{max}$ vs. mass, for different values of local density.
      }
       \begin{center}
       \begin{tabular}{ccc}
       \hline\hline
      Local density parameter ($\rho_n$)& $\gamma$ & $m_0$\\
       \hline\hline
      All $\rho_n$ values & 0.295 & $1.29\times10^{5}$\\
      $\rho_n > 2\bar{\rho_n}$ & 0.281 & $4.22\times10^{4}$\\ 
      $\bar{\rho_n}/2 < rho_n < 2\bar{\rho_n}$ & 0.404 & $4.85\times10^{7}$
      \\
      $\rho_n < \bar{\rho_n}/2$ & 0.442 & $1.36\times10^{8}$\\
       \hline\hline
       \end{tabular}
       \end{center}
       \label{tab:samp2}
\end{table}

\begin{figure}
       \includegraphics[width=0.47\textwidth]{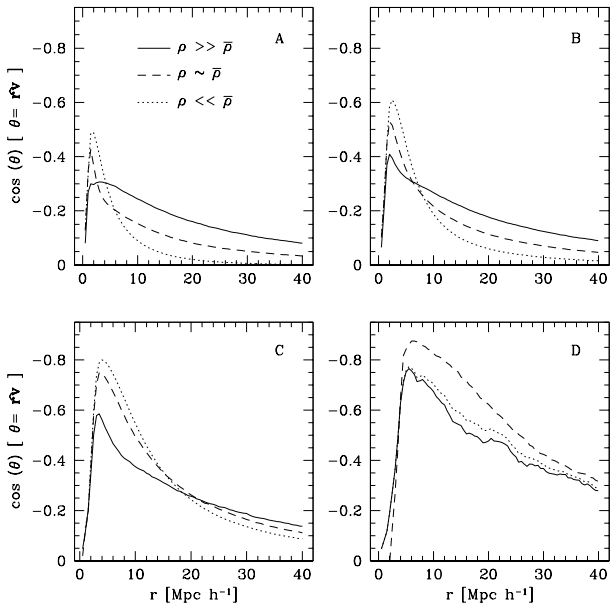}
       \caption{Dependence of $\left<cos(\theta)\right>$ on local
density for our four mass groups samples. Panel A: 
$\left<cos(\theta)\right>$ for groups in mass sample S1, computed for 
particles in regions with $\rho >> \bar{\rho}$ (solid line), 
$\rho \sim \bar{\rho}$ (dashed line) and $\rho << \bar{\rho}$ (dotted line). 
Panel B: same as A, for groups in sample S2. Panel C: r groups in 
sample S3. Panel D: groups in sample S4.}
       \label{fig6}
\end{figure}

In order to get a deeper insight on the flow patterns in the 
outskirts of groups, we also study the dependence of the {\it{infall-angle}} 
on density; we compute the average $\left<cos(\theta)\right>$ for particles 
characterised by a local density $\rho_{n}$.
Figure \ref{fig6} shows these results for the 
mass samples defined in Table \ref{tab:samp} (panels A, B, 
C and D for samples S1, S2, S3 and S4 respectively) for particles in
high density environments, $\rho >> \bar{\rho}$ (solid lines); 
intermediate densities, $\rho \sim \bar{\rho}$ (dashed line) and 
low densities, $\rho << \bar{\rho}$ (dotted line). Each group sample shows
diferent features in their alignment profiles including different maximum
alignments; we notice that for higher mass groups the maximum alignment is 
more significant, as can be seen in Figure \ref{fig5}. For the least massive 
groups (sample S1 in panel A) we obtain a maximum alignment 
of $\left<cos(\theta)\right> \sim -0.5$ at
$r \sim 3h^{-1}$ Mpc for low density particles, and a lower maximum 
alignment for particles in high density regions, 
$\left<cos(\theta)\right> \sim -0.25$ at $r \sim 4h^{-1}$ Mpc.  
However, at larger scales, $r > 6h^{-1}$ Mpc, high-density particles are 
better aligned than low-density particles. Samples S2 and S3 
(intermediate masses) show a similar behaviour: the maximum alignment of
low-density particles around groups in sample S2 reaches
$\left<cos(\theta)\right> \sim -0.6$, whereas the high-density particles 
reach $\left<cos(\theta)\right> \sim -0.4$, both at $r > 8h^{-1}$ Mpc.  
As can be seen, the peculiar velocities of higher-density particles
are more aligned with the direction to the centre of the halo.  
In the case of group sample S3 the particles in low density regions have a 
maximum alignment $\left<cos(\theta)\right> \sim -0.8$ whereas high-density 
particles show a maximum alignment $cos(\theta) \sim -0.6$.  Again, at scales 
$r > 20h^{-1}$ Mpc, higher
density particles are more aligned. In group sample S4 it can be observed that  
particles in low and high density  environments show
the same maximum alignment $cos(\theta) \sim -0.8$ at 
$r \sim 6h^{-1}$ Mpc, while particles with intermediate densities are 
more strongly aligned, with $cos(\theta) \sim -0.9$. 
For this particular group sample, we do not 
find a significant difference in the alignment of high, medium, and low density 
particles.  
Table \ref{tab:samp3} shows the parameter values for the fits
to $cos(\theta)_{max}$ vs. mass for the different ranges of local densities
and halo mass
considered here.  As can be seen, both the slope $\gamma$ and $c_0$ are 
larger for higher densities.  This indicates that particles in denser regions
are moving in a direction better aligned to the center of the groups.
As in the case of the maximum infall velocity, it can also be seen that
the parameter fits resulting from all the particles surrounding groups is
similar to the fit from the particles in the densest regions, indicating
that the latter dominate the alignment signal.

\begin{table}
 \caption{Values obtained for the parameters $\gamma$ and $C_0$ in the fit of
$cos(\theta)_{max}$ vs. mass, considering different ranges of local density.}
 \begin{center}
 \begin{tabular}{ccc}
 \hline\hline
Local density ($\rho_n$) & $\gamma$ & $C_0$\\
 \hline\hline
All $\rho_n$ values & 0.176 & $5.18\times10^{10}$\\
$\rho_n > 2\bar{\rho_n}$ & 0.178 & $6.16\times10^{10}$\\ 
$\bar{\rho_n}/2 < rho_n < 2\bar{\rho_n}$ & 0.159 & $2.66\times10^{9}$
\\
$\rho_n < \bar{\rho_n}/2$ & 0.157 & $1.05\times10^{9}$\\
 \hline\hline
 \end{tabular}
 \end{center}
 \label{tab:samp3}
\end{table}

\begin{figure}
       \includegraphics[width=0.47\textwidth]{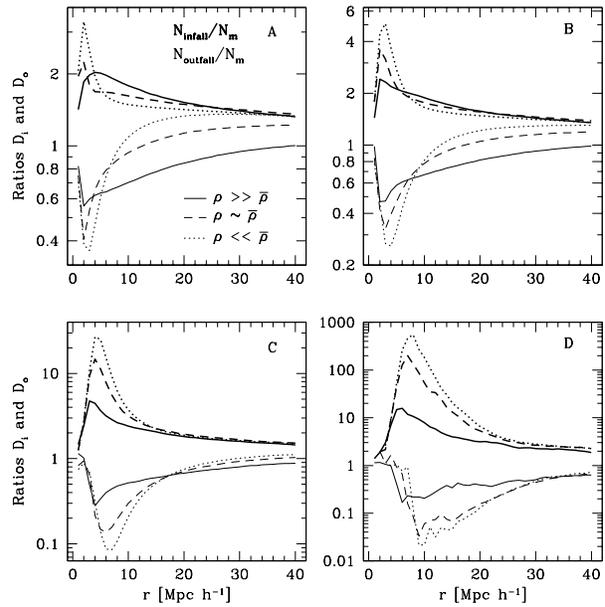}
       \caption{Dependence of $N_{outflow}/N_{m}$ (black) and 
$N_{infall}/N_{m}$ (gray) with the local density of dark-matter particles, for 
four subsamples of halos (panels A, B, C and D for samples S1, S2, S3 and S4, 
respectively). Results for $D_i$ and $D_o$ for particles characterised by
$\rho/\bar{\rho} > 2$ are shown in solid lines, whereas dashed
lines shows the results for $\rho \sim \bar{\rho}$, and dotted lines the results
for $\rho/\bar{\rho} < 1/2$.}
       \label{fig8}
\end{figure}

We now search for possible dependencies of the fraction of infalling
and outflowing particles, $D_{o}$ and $D_{i}$ defined above, on the 
local density of dark-matter particles.  Figure \ref{fig8} presents 
these results for our four group samples (panels A, B, C 
and D for samples S1, S2, S3 and S4, respectively) for particles in 
low- (dotted line), intermediate- (dashed line) and 
high-density (solid line) regions. As can be seen, the ratio $D_{i}$ is 
larger for low-density particles at small separations, whereas at larger 
scales, the trend is similar for particles in both, high and low density 
regions. Inspection of panel D, indicates that an important difference 
of three orders of magnitude can be achieved between
the number of infalling to outflowing particles for high density regions.  
Note that this maximum occurs at a scale $\sim 50\%$ larger than that 
in low density regions. In high density regions, however, 
this difference is not so important.
It should also be noticed that, although not detected in Figure 6 for
the overall ratios of infalling and outflowing particles, particles
in the lower density regions show a minimum in $D_i$ at slightly (marginally)
larger distances from the halo centres than the corresponding maximum in $D_o$.

\section{Conclusions}

We now summarise the main conclusions we arrived at during the series of 
analyses performed on the VLS simulations.  We studied the peculiar velocity 
field around dark-matter haloes of varying masses, considering infall 
velocities, the {\it infall angle} with respect to the halo centres, and the 
fractions of infalling and outflowing particles.  We also studied variations 
in these quantities as the local dark-matter particle density varies.

We now list our main results and conclusions:

\begin{itemize}
\item The maximum infall velocity around the 
most massive groups identified in the numerical simulation
($\sim 10^{15}h^{-1} M_{\odot}$), is $v_{inf}\sim 900$ km/s. The results for 
the lowest mass group sample ($\sim 10^{12}h^{-1}M_{\odot}$) indicate a 
maximum infall velocity of $\sim 220$ km/s, decreasing very slowly at 
larger distances from the halo centres.

\item The maximum infall velocity occurs in a narrow range of scales,
$2h^{-1}$ Mpc $< r_{max} < 6h^{-1}$ Mpc even when considering a broad range 
of group masses ($10^{12}h^{-1}M_{\odot}$ to $10^{15}h^{-1}M_{\odot}$).

\item The exponential cutoff linear model analysed by Croft et al., works 
better for the most massive groups at small scales. It does not 
reproduce the infall pattern for groups samples with 
$M < 10^{14}h^{-1}M_{\odot}$, neither the large scale infall signature 
onto massive systems in any of our subsamples S1 to S4.

\item We find a power law relation between the maximum infall velocity 
and mass. The Croft et al. model with an exponential overdensity cutoff 
provides a good fit for $\delta_{c} \simeq 45$.

\item Large mass groups show a stronger alignment between the 
dark-matter particle peculiar velocity and its cluster-centric position 
than do low mass groups. Regardless of the halo mass, this maximum 
alignment occurs at intermediate scales, 
$2h^{-1}$ Mpc $< r_{max} < 8h^{-1}$ Mpc.

\item There are remarkable differences between the dynamical behaviour of low-
and high-density particles.  In general, higher density particles exhibit
higher infall velocities and lower alignment than particles in lower density
environments.  The difference in infall is less evident when considering
high mass haloes, and the alignment becomes similar for high and low density
particles for larger distances to halo centres.  However, at large enough 
distances, high-density particles become even more aligned than lower 
density particles, specially for low mass haloes.

\end{itemize}

We have also considered the possibility that substructure could affect
our analysis. According to the results of Aubert, Pichon \& Colombi (2004) 
the infall pattern of the largest halos is affected by substructure only 
at small scales ($r<2 Mpc$). Since the maxima occurs well beyond this scale, 
we are confident that our results are not likely to change under the presence 
of substructure.

In spite of the fact that mass flows toward clusters in an anisotropic
fashion preferentially through the filamentary structure, the results
provided give useful mean values of velocity flows which can be used to
asses the systematic infall onto growing structures.
The particular analysis performed provide direct measures of the maximum
infall pattern and halo mass dependence and can be used to estimate gas
dynamical processes such as ram pressure stripping that may initiate
galaxy transformation in the outskirts of clusters.

\section*{Acknowledgments}
MP acknowledges receipt of a CONICET fellowship.
NDP was supported by a Proyecto FONDECYT Postdoctoral 3040031.  This work was 
supported in part by the "ESO Center for Astrophysics" at Catolica 
and by the European Commission's ALFA-II programme through its funding of 
the Latin-american European Network for Astrophysics and Cosmology, LENAC.
The simulations in this paper were carried out by the Virgo Supercomputing 
Consortium using computers based at the Computing Centre of the Max-Planck 
Society in Garching and at the Edinburgh Parallel Computing Centre. The data 
are publicly available at http://www.mpa-garching.mpg.de/NumCos. This research 
has made use of NASA's Astrophysics Data System.


\end{document}